# High-fidelity far-field microscopy at λ/8 resolution

*Ning Xu, Guoxuan Liu, and Qiaofeng Tan\**

N. Xu, Dr. G. Liu, and Prof. Q. Tan
State Key Laboratory of Precision Measurement Technology and Instruments, Department of
Precision Instrument, Tsinghua University, Beijing 100084, China
E-mail: tanqf@mail.tsinghua.edu.cn

**Abstract**

The emergence of far-field super-resolution microscopy has rejuvenated the possibility for nanoscale imaging. Approaches to far-field super-resolution that utilize point scanning often depends on spatially reducing the size of the focused spot. However, the focused spot always achieves high resolution at the expense of extremely low light efficiency for the probing mainlobe and high-intensity sidelobes, which limits the applications in nanoscale imaging and might cause misinterpretation of samples. Here we report a sharp probing spot with the diffraction efficiency of 3.76% at the resolution of 38% of the Airy spot size assisted by the two-dimensional multi-level diffractive optical element (DOE) experimentally. The diffraction efficiency of DOE is improved by at least two orders of magnitude at the same resolution by breaking the limitation of circular 0-π binary structure superoscillatory lens. To eliminate the influence of the high-intensity sidelobes, high-fidelity images are reconstructed based on the modified deconvolution algorithm by virtue of the prior-knowledge. Finally, high-fidelity far-field microscopy (HiFi-FM) is constructed and experimental results show that HiFi-FM allows the resolution of spatially complex samples better than 69 nm while acquiring high fidelity.

## 1. Introduction

Classical criterion for microscopic system reveals the basic resolution limited by the wavelength of beam λ and the numerical aperture (NA) restricts lateral spatial resolution to ca. 200nm [1]. To observe nanoscale objects, researchers in optics spare no effort to pursue far-field super-resolution imaging by improving spatial frequency response of the optical transfer function (OTF) [2, 3] or increasing the cut-off of the OTF [4]. For example, structured illumination microscopy (SIM) [5] increases the cut-off of the OTF and confocal microscopy (CM) [6, 7] improves the spatial frequency response of the OTF. Decreasing the size of the focused spot for illumination is one of the most promising tools for improving the spatial frequency response of the OTF [8, 9].

As originally proposed by Toraldo di Francia [10], focusing beyond the Abbe-Rayleigh diffraction limit can be achieved by a pupil filtering technique [8, 11, 12]. Notably, the superoscillatory lens [13-16] and diffractive optical element (DOE) [17-19] have been studied to generate a tightly focused spot, and the image of the sample is reconstructed in the point-by-point manner, in a certain case, resolution up to $\lambda/6$ [20]. The ongoing challenge towards high-resolution imaging still calls for an enhancement in spatial resolution in practice.

Although these advancements have been made, extremely low diffraction efficiency of the probing spot and high-intensity halo restrict the application stick out. Typically, the efficiency of the reported circular $0$-$\pi$ binary structure superoscillatory lens for the resolution of 30% of the Airy spot is far less than 0.01% [15, 21]. It has been evident that the focused spot inevitably contains the sidelobes when the focused spot size is smaller than the criterion of $0.38\lambda/NA$ by a $0$-$\pi$ superoscillatory lens [22]. Besides, the fidelity of the super-resolution imaging is also challenged due to the influence of the high-intensity halo. In a point-by-point manner, lots of researchers focused on how to decrease the width of the mainlobe to improve the resolution [8, 17, 19, 21-24]. However, the intensities and distributions of sidelobes of the focused spot have an effect on the final reconstructed super-resolution images. Especially, biological structures discovered using super-resolution techniques need to be interpreted with high-fidelity (HiFi) to avoid misinterpretation. Several researchers have been pursued to exploit reconstruction algorithms, containing iterative deconvolution [25-27] and engineering the point spread function (PSF) into an ideal form [28]. However, reconstruction algorithms of point-by-point manner have rarely been exploited to achieve HiFi super-resolution images.

Here, we propose and experimentally demonstrate a high-fidelity far-field microscopy (HiFi-FM) that can reconstruct super-resolution complex images with a spatial resolution at $\lambda/8$. In contrast to a circular binary superoscillatory lens, a two-dimensional multi-level DOE improves the efficiency of the probing spot. We generate a sharp probing spot with the efficiency of 3.76% at the resolution of 38% of the Airy spot size, and the efficiency at least increases by two orders of magnitude than that realized by superoscillatory lens at the same resolution. However, the sidelobes still have an obvious effect on the reconstructed super-resolution images, and then a reconstruction algorithm based on the modified deconvolution algorithm with the prior-knowledge is proposed. HiFi-FM can achieve spatial resolution of 69 nm for spatially complex samples with high-fidelity, and the results are compared with those of widefield and non-fidelity (NoFi) reconstruction.

## 2. Principles

### 2.1. HiFi-FM Theory and Demonstration

The idea of HiFi-FM is sketched in Fig. 1a. The system consists of two components: the illumination system (blue ray, see Fig. S1 of Supplement 1) and the imaging system (green ray, Fig. S1). Specifically, the focused spot produced by the DOE ($xy$-plane, DOE plane) with illumination objective is diffracted in the sample ($uv$-plane, Fourier plane). The sample is scanned by the tightly focused spot, termed as the probing spot, with the requirement of Nyquist-Shannon sampling to obtain super-resolution information, and the other 8-spots around are termed as the constrained spots to assist in generating such a probing spot. There are clearly distinguishable dark regions between the probing spot and the constrained spots. The intensity values $I(\xi, \eta)$ (fluorescence or nonfluorescence from the sample) on the detector can be obtained by the imaging system, and the value $I(0,0)$ corresponds to the center of the probing spot. The discretized values of $I_{(m, n)}(0,0)$ can be obtained during scanning and are used to generate the super-resolution image, here subscript $(m, n)$ shows the scanning position of the probing spot. For simplification, $I_{(m, n)}(0,0)$ is written as $I(m, n)$. The "Experimental Set-up" section and Fig. S1 of Supplement 1 present a complete description of the experimental geometry.

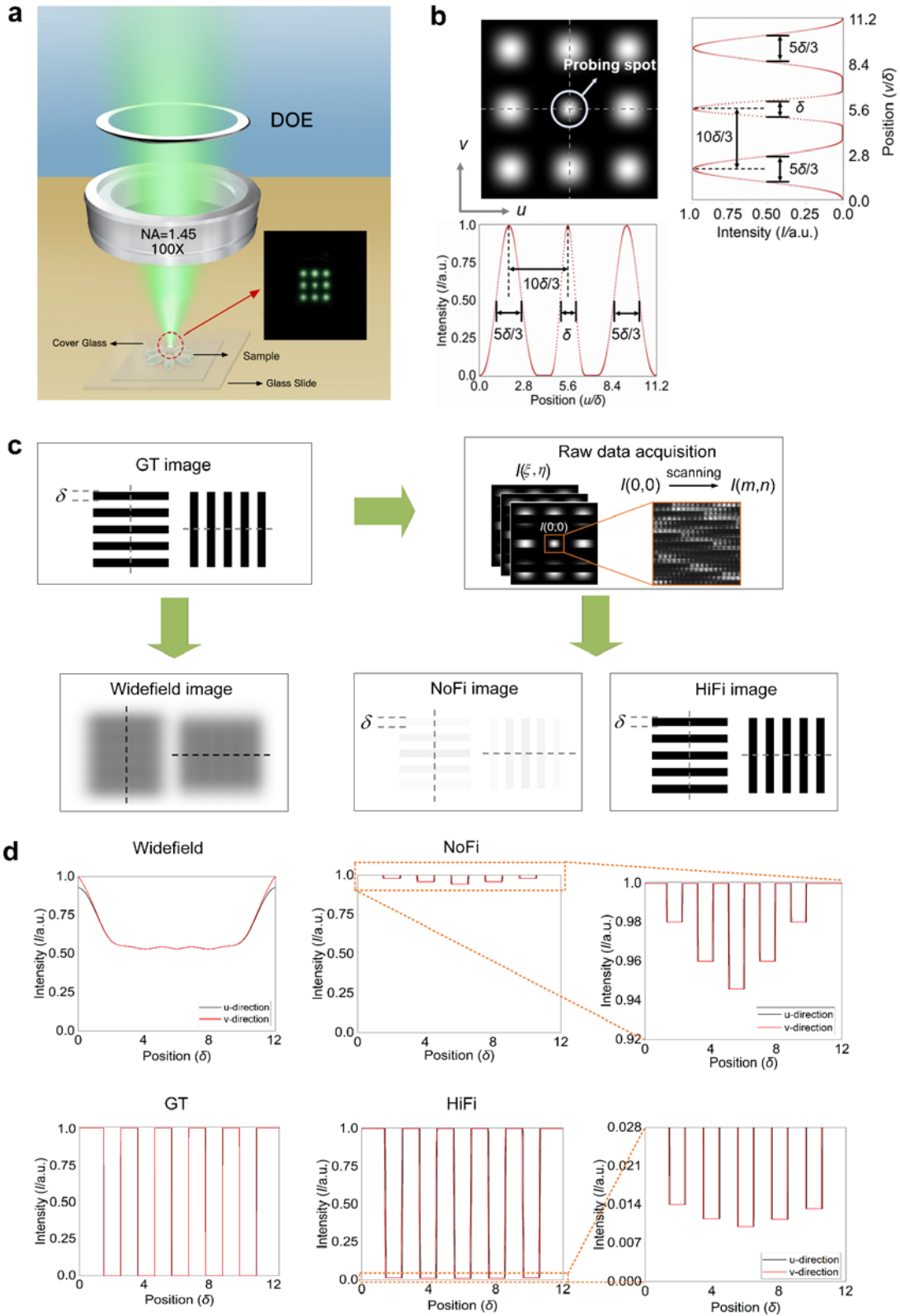

**Figure 1. Underlying principle of HiFi-FM. a** Schematic of the DOE integrated with an inverted microscope. **b** The full width with half maximum (FWHM) of probing spot and the around constrained spots are $\delta$ and $5\delta/3$, respectively. **c** Flowchart of the widefield image, NoFi image, and HiFi image with the linewidth of $\delta$ of the positive resolution test target. Ground-truth (GT) is the positive resolution test target with the linewidth of $\delta$. **d** The black and red lines are the corresponding lines with the images of the widefield, NoFi, HiFi, and GT from **c** in the $u$-direction and $v$-direction.

Super-resolution information may not be directly obtained by the imaging system, albeit the illumination system enables a probing spot to acquire the super-resolution information from the sample. Consequently, it is a critical point to find a relationship to reconstruct the super-resolution information of the sample. The intensity of HiFi-FM on the detector can be analytically expressed as [29]

$$I\left(\xi,\eta\right)=\iint p_{\mathrm{illu}}\left(u,v\right)t\left(u,v\right)p_{\mathrm{imag}}\left(\xi-u,\eta-v\right)dudv \,, \tag{1}$$

where $p_{\mathrm{illu}}(u, v)$ and $p_{\mathrm{imag}}(\xi, \eta)$ are respectively the PSFs of the illumination system and imaging system, and $t(u, v)$ is the intensity transmittance function of the sample. Here, the PSF of illumination system $p_{\mathrm{illu}}(u, v)$ is the key factor to realize super-resolution imaging via the point-by-point manner. Supposing the probing spot and 8-constrained spots are shown in Fig. 1b, with the sampling interval of $\delta$, the desired probing spot is defined within the window $[-\delta/2, \delta/2]$ and

$$p_{\mathrm{illu}}\left(u,v\right)=\begin{cases}p_{\mathrm{ps}}\left(u,v\right) & |u|\le \delta/2,\ |v|\le \delta/2 \\ p_{\mathrm{other}}\left(u,v\right) & \mathrm{else}\end{cases}, \tag{2}$$

where $p_{\mathrm{ps}}(u, v)$ and $p_{\mathrm{other}}(u, v)$ represent the PSF values of the probing spot and other area (sidelobes including the 8-constrained spots), respectively. The observed intensity of the probing spot $I(0,0)$ can be expressed as

$$I\left(0,0\right)=\iint p_{\mathrm{illu}}\left(u,v\right)t\left(u,v\right)p_{\mathrm{imag}}\left(-u,-v\right)dudv \,. \tag{3}$$

For the sparse sample on the dark-background, especially for a single square bright spot with size $\delta$ and transmittance $t(0,0)$ on dark-background (scan the sample), the transmittance function can be written as

$$t\left(u,v\right)=\begin{cases}t\left(0,0\right) & |u|\le \delta/2,\ |v|\le \delta/2 \\ 0 & \mathrm{else}\end{cases}, \tag{4}$$

and $t(0,0)$ and $I(0,0)$ have an object-image relation realized by the imaging objective. In this case, Eq. (3) can be written as

$$I\left(0,0\right)=t\left(0,0\right)\cdot \int_{-\delta/2}^{+\delta/2}\int_{-\delta/2}^{+\delta/2}p_{\mathrm{ps}}\left(u,v\right)p_{\mathrm{imag}}\left(-u,-v\right)dudv \,. \tag{5}$$

From the observed intensity $I(0,0)$ of the probing spot and the ascertainable $p_{\mathrm{ps}}(u, v)$ and $p_{\mathrm{imag}}(-u, -v)$, the transmittance $t(0,0)$ can be directly obtained by Eq. (5), and the spatial resolution of the system is decided by $\delta$ of the probing spot.

In contrast to sparse sample on the dark-background, complicated transmittance function as biological sample, to a large extent, affects the authenticity of result due to the convolution effect of sidelobes. Equation (3) can be further expressed as

$$I(0,0) = \sum_i \sum_j P_{\text{illu}}(i,j) T(i,j) P_{\text{imag}}(-i,-j) \ . \tag{6}$$

where $P_{\text{illu}}(i,j)$, $T(i,j)$, and $P_{\text{imag}}(i,j)$ represent the discretized parameters of the PSF of the illumination system, the transmittance function of the sample, and the PSF of the imaging system, respectively.

According to Eq. (6), the relationship between $I(0,0)$ and the transmittance function of the sample $T(0,0)$ can be bridged with the sampling interval of $\delta$. The observed intensities $I(m, n)$ can be obtained during scanning, which can be further expressed as

$$I(m,n) = \sum_{i=-L}^{+L} \sum_{j=-L}^{+L} P_{\text{illu}}(m+i,n+j) T(m+i,n+j) P_{\text{imag}}(m-i,n-j) \ . \tag{7}$$

where $P_{\text{illu}}(m, n)$, $T(m, n)$, and $P_{\text{imag}}(m, n)$ respectively represent the discretized parameters of the PSF of the illumination system, the transmittance function of the sample, and the PSF of the imaging system during scanning. The number of discretized points $L \times L$ are decided by the size of the illumination spots, and the discretized $P_{\text{illu}}(m, n)$ and $P_{\text{imag}}(m, n)$ can be determined by the parameters of the system.

The FWHMs of the probing spot and the constrained spots are respectively $\delta$ and $5\delta/3$, and the corresponding PSF is shown in Fig. 1b. The observed image on the detector in Fig. 2c is blurred after convolution of the PSF of the illumination system in Fig. 2a and the PSF of the imaging system in Fig. 2b. It can be seen from Eq. (7) and Fig. 2c that $I(m, n)$ is influenced by the intensities and distributions of sidelobes, which affects the authenticity of the reconstruction images.

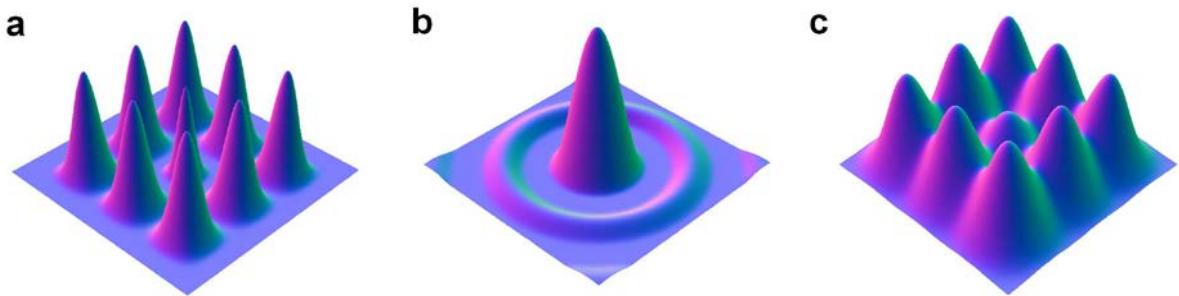

**Figure 2. PSF analysis of HiFi-FM. a** PSF of the illumination system. **b** PSF of the imaging system, respectively. **c** The effective PSF of the convolution of the PSF of the illumination system and the PSF of the imaging system. Images size: $12\delta \times 12\delta$, $\delta = 0.183\lambda/\text{NA}$.

## 2.2. Fidelity characterization with standard sample

To present the performance of HiFi-FM, we have adopted the size of probing spot and constrained spots with the FWHM of $\delta$ and $5\delta/3$, as shown in Fig. 1b. The positive resolution test target with the linewidth of $\delta$ is used as the sample to assess the performance by the probing spot via the point-by-point manner. Considering the convenience for practical application, the NAs of the imaging objective equals that of the illumination objective, and as a consequence that an inverted microscope can be used to construct the HiFi-FM. The superimposed observed intensity of the probing spot $I(m, n)$ is shown in raw data acquisition section of Fig. 1c. On account of the neglect of the sidelobes, the discretized transmittance function of the sample $T(m, n)$ can be expressed as Eq. (10). Interestingly, we have found that the linewidth of $\delta$ can be resolved, however, the transmittance function of the sample deviates greatly from its ground truth due to the sidelobe especially for dark objects with bright-background, as shown in NoFi image of Figs. 1c and 1d.

To ensure the fidelity of image, assisting with the prior-knowledge, super-resolution images with HiFi are reconstructed without increasing the complexity of the system. Notably, the resolution of the HiFi-FM is determined by the probing spot while the modified HiFi algorithm does not influence the resolution of reconstruction image. The flowchart and plots of the widefield, NoFi, HiFi, and GT with the linewidth of $\delta$ of the resolution test target are presented in Figs. 1c and 1d.

Figure 3 shows the contrasts of different images from a positive resolution test target sample. HiFi image of the positive resolution test target guarantees the fidelity while ensuring the resolution (Fig. 1c and Fig. 3a). With HiFi-FM, the fidelity of the image is to a great extent improved. The contrast and structural similarity index measure (SSIM) of images to the ground-truth image were respectively improved from 0.027 to 0.99 and 0.14 to 0.98 by HiFi-FM than that by NoFi-FM (Fig. 3a and Fig. S5). For the sample on the bright-background, the differences of fidelity are usually more obvious than those on the dark-background. The calculated contrast remained above 95% to the ground-truth image, even with different reconstruction parameters ($\alpha_1$, $\alpha_2$, and $\alpha_3$), as shown in Fig. 3a. This suggests that super-resolution images by the HiFi algorithm are not sensitive to the reconstruction parameters. Thus, routine adjustment is not required for most cases, which considerably facilitates the application of HiFi-FM as a daily imaging tool.

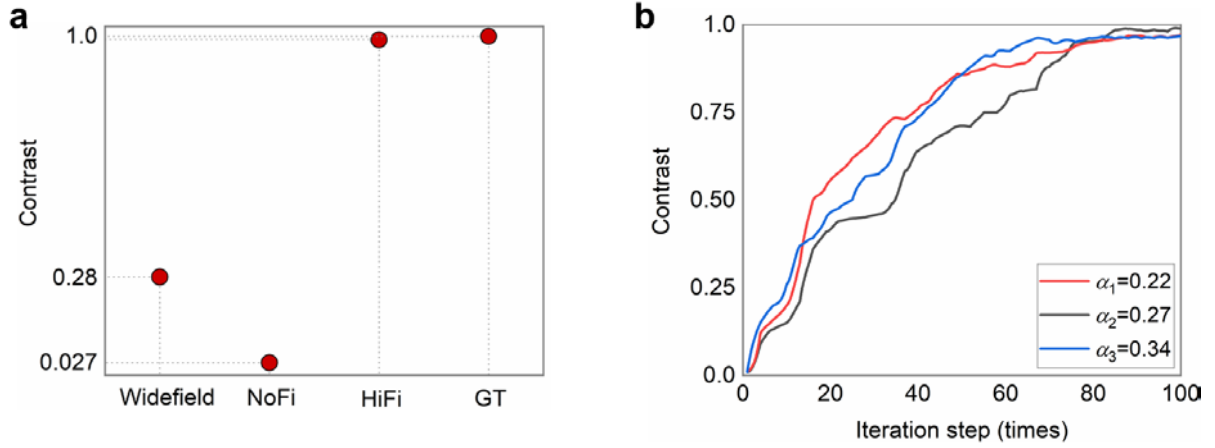

**Figure 3. Quantitative characterization of the fidelity of the positive resolution test target reconstruction. a** Contrast plot of the HiFi images with different reconstruction parameters (α=0.22, 0.27, and 0.34). **b** Comparison of the contrast of the images in Figs. 1c and 1d obtained by widefield, NoFi, HiFi, and GT.

## 3. Methods

### 3.1. Generation of super-resolution probing spot

To obtain a higher resolution with practical efficiency, 8-constrained super-resolution spots around the probing spot are firstly adopted, as shown in Fig. 1b, as the ideal intensity distribution $I_{\text{ideal}}(u,v)$ in the Fourier plane of the illumination objective for the first time. In contrast with the iterative Fourier transform algorithm (IFTA) [30], a more precise initial phase and an effective constraint in the Fourier plane are proposed to design the phase distribution. Here, the initial phase distribution in the Fourier plane is estimated as

$$\psi^{(1)}(u,v) = \cos\left[uv\cos(u)\cos(v) + \beta(u^2+v^2)\right],\tag{8}$$

where $\beta$ is a positive constant. The former item will produce super-resolution probing spot and constrained spots, and the latter defocused item will make the DOE plane be closer to pure-phase distribution. In addition, the constraint in the Fourier plane for the propagated field is combined with the target intensity distribution expressed as [31]

$$U^{(i+1)}(u,v) = \left\{\Gamma\sqrt{I_{\text{ideal}}(u,v)} + (1-\Gamma)\mathcal{T}^{(i)}(u,v)\right\}\exp\left\{j\psi^{(i)}(u,v)\right\},\tag{9}$$

where $\Gamma$ controls the relative amplitude field distribution within 0 to 1, $i$ is the iterative number, $\mathcal{T}^{(i)}(u,v)$ and $\psi^{(i)}(u,v)$ are respectively the amplitude and phase distribution in the Fourier plane. The light field $U^{(i+1)}(u,v)$ in the Fourier plane is inverse Fourier transformed to calculate the light field $E^{(i+1)}(x,y)$ in the DOE plane.

By tradeoff the resolution and efficiency with HiFi result in practice, we have validated the proposed method to respectively generate a probing spot and constrained spots with the FWHM of 30% and 50% of the Airy spot under the NA=1.45, adopted $i$=500, $\beta$=0.03, and $\Gamma$=0.27. The phase distribution and numerical simulation results of the probing spot and constrained spots are shown in Figs. 4a and 4b, respectively. The experimental carrier frequency with tilt factor in Fig. 4a is to avoid influence from the zeroth-order beam. The FWHM of the probing spot is theoretically narrowed to 36% compared with the FWHM of the Airy spot. The measured FWHM of the probing spot is experimentally narrowed to 38% of the Airy spot, a resolution of 65.2 nm (~1/8 of the incident wavelength) with an experimental efficiency up to 3.76% is achieved, as shown in Figs. 4c and 4d. The distributions of the probing spot and the constrained spots in Figs. 4b and 4c are different from the ideal spot in Fig. 1b, $P_{\text{illu}}(i, j)$ used in the HiFi image reconstruction algorithm needs to determine according to the experimental result in Fig. 4c.

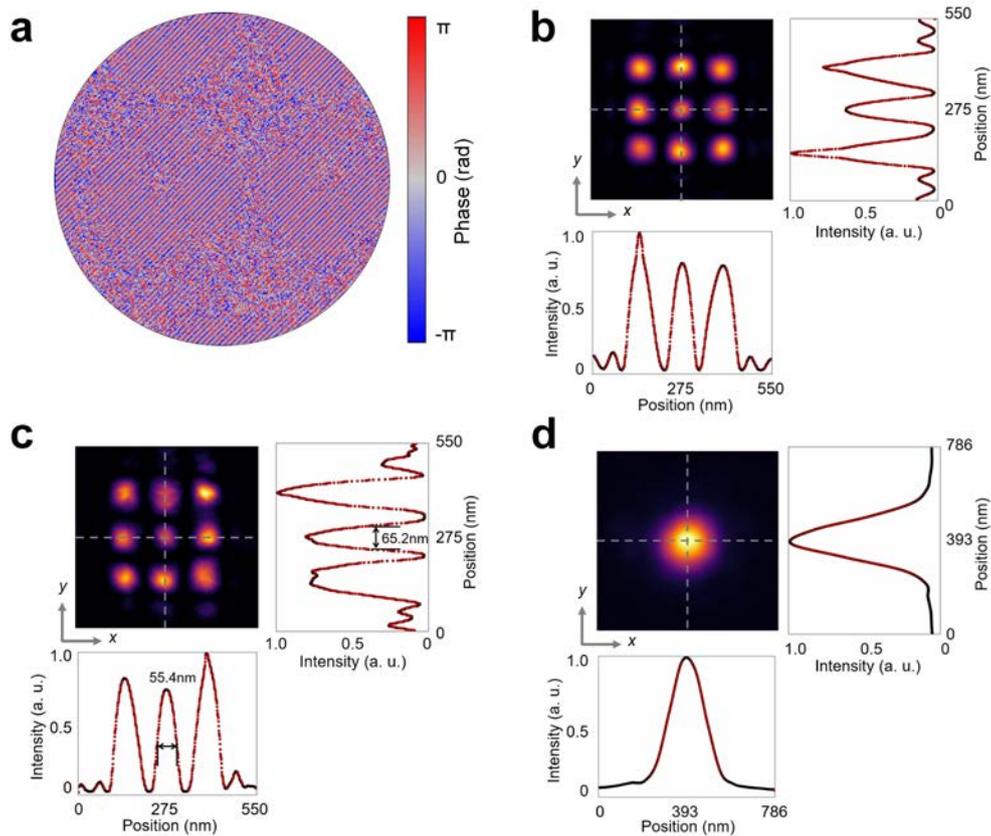

**Figure 4. Generated probing spot of HiFi-FM in numerical simulation and experimental results by the proposed method. a** Phase distribution for generating a probing spot and constraint spots. **b** and **c** are the numerical simulation and experimental results of the probing spot and constrained spots, respectively. **d** Experimental result of Airy spot. Objective lens: 100×/1.45 NA.

### 3.2. HiFi reconstruction algorithm

To reconstruct the discretized transmittance function of the sample $T(m, n)$, the general method, termed as NoFi image reconstruction algorithm, is approximately regarded as [32]

$$T_{\text{NoFi}}(m,n) = \frac{I(m,n)}{P_{\text{illu}}(m,n) P_{\text{imag}}(m,n)} \ . \tag{10}$$

The influence of the sidelobes is not considered in the Eq. (10), which only can be applied to the sparse sample on the dark-background.

For most samples, every term of $P_{\text{illu}}(m, n) \cdot P_{\text{imag}}(m, n)$ in Eq. (7) should be considered. The HiFi of the transmittance function of the sample $T(m, n)$ is expected to reconstruct from $I(m, n)$, Eq. (7) can be further expressed as

$$\begin{bmatrix} T(m-L,n-L) \\ \vdots \\ T(m,n) \\ \vdots \\ T(m+L,n+L) \end{bmatrix}^T \begin{bmatrix} P_{\text{illu}}(m-L,n-L) \cdot P_{\text{imag}}(m+L,n+L) \\ \vdots \\ P_{\text{illu}}(m,n) \cdot P_{\text{imag}}(m,n) \\ \vdots \\ P_{\text{illu}}(m+L,n+L) \cdot P_{\text{imag}}(m-L,n-L) \end{bmatrix} = I(m,n) \ . \tag{11}$$

Unfortunately, an indefinite equation of Eq. (11) cannot be solved directly. Hence, we put forward a modified deconvolution algorithm to make use of the prior-knowledge that reconstructs super-resolution image with HiFi.

Most previous works on fidelity-improved imaging used deconvolution to reconstruct HiFi image [25-28, 33-35]. While this has been demonstrated to achieve HiFi imaging, it has also encountered problems such as the input estimated value and criteria heavily decided the performance of results [36, 37]. Hence, we propose a modified Richardson-Lucy deconvolution algorithm [33, 34] by virtue of the prior-knowledge to reconstruct HiFi results. This modified algorithm departs from the conventional deconvolution for which the constrained spots and known-transmittance area are respectively adopted as the input value and criteria of the results.

To generate a precise input value, we pick off the transmittance $T_p$ at the positions where the probing spot does not illuminate the sample while the sidelobes do during the scanning. Under such conditions, the transmittance of the probing spot is actually zero, however, $T_p$ is not zero due to the effect of convolution. Such $T_p$ can be regarded as the prior-knowledge. The input image $T_{\text{in}}(m, n)$ can be expressed as

$$T_{\text{in}}(m,n) = \alpha \left[ T_{\text{NoFi}}(m,n) - \langle T_p \rangle \right] + (1-\alpha) T_{\text{WF}}(m,n) \ . \tag{12}$$

where $T_{NoFi}(m, n)$ can be calculated by Eq. (10), $\alpha \in (0, 1)$, $T_{WF}(m, n)$ and $<T_p>$ are respectively the transmittance distribution on the widefield and the average transmittance at the picking off positions.

Moreover, we utilize the known-transmittance area on the sample to achieve a better result with reasonable criteria. If the total transmittance of the known-transmittance area is closer to the default value, the result is closer to fidelity. More detailed information on HiFi reconstruction is presented in Note 1 and Fig. S3 of Supplement 1.

### 3.3. Quantification of the fidelity of reconstruction images

To quantitatively evaluate the fidelity of images reconstructed by HiFi-FM, the positive resolution test target of known real structure was employed as a standard sample for imaging (Figs. 1c and 1d). Because structures of the samples were known, raw data with enough exposure time were collected and super-resolution images with HiFi were reconstructed thereafter. The influence of the sidelobes and speckles in the NoFi image was to a great extent eliminated by proposed HiFi reconstruction algorithm, and reconstructed images were obtained as the ground-truth image. Contrast maps and corresponding contrast values were displayed to evaluate the fidelity of reconstruction algorithms (Fig. 3a). The Michelson contrast was defined as [38, 39]

$$\text{contrast} = \frac{I_{\max} - I_{\min}}{I_{\max} + I_{\min}} \quad, \tag{13}$$

where $I_{\max}$ and $I_{\min}$ are respectively the maximum and minimum intensity. Furthermore, the structural similarity index measure (SSIM) [40] was used to quantitatively evaluate the fidelity of images, defined as

$$\text{SSIM}\left(I_{RI}, I_{GT}\right) = \frac{\left(2\mu_{RI}\mu_{GT} + C_1\right)\left(2\sigma_{RI,GT} + C_2\right)}{\left(\mu_{RI}^2 + \mu_{GT}^2 + C_1\right)\left(\sigma_{RI}^2 + \sigma_{GT}^2 + C_2\right)} \quad, \tag{14}$$

where $I_{RI}$ and $I_{GT}$ are the intensity of reconstruction image and ground-truth image, respectively; $\mu_{RI}$ and $\mu_{GT}$ are the mean values of $I_{RI}$ and $I_{GT}$ in the same position, respectively; $\sigma_{RI}$ and $\sigma_{GT}$ are the standard deviations of $I_{RI}$ and $I_{GT}$, respectively; and $I_{RI,GT}$ is the covariance between images $I_{RI}$ and $I_{GT}$; $C_1$ and $C_2$ are the constant parameters. A higher SSIM means that the region of these two images is more similar.

### 3.4. Experimental set-up

We carried out experiments with incident linearly polarized wavelength at 488 nm, which were performed from a laser (Sapphire 488, Coherent) with a spatial filter consisted of a lens L1 with the focal length $f_1$=100 mm and a 5 μm pinhole (M-900, Newport). The output energy was ~50mW, which could be modulated by an attenuator. A reflective phase-only spatial light modulator (SLM; PLUTO, Holoeye) was used to upload the phase pattern as a DOE, where the pixel size is 8 μm×8 μm. The pixel number of SLM is 1080×1920, and the circular area with a diameter of 500 pixels in the center was used. The aperture was utilized to ensure the used area of the SLM was circular with a diameter of 4 mm. Parallel beams were separated into two paths with orthogonal directions by a beam splitter (BS) cube (CCM1-BS013/M, Thorlabs). The beam modulated by SLM was expanded by the telescope to match the area of the entrance pupil of the illuminating objective. Two lenses were placed in a 4$f$ configuration to image the phase distribution on the objective pupil to fill the entrance pupil of the microscope objective, where $f_2$=400 mm and $f_3$=500 mm. The polarizer was also used to guarantee the incident beam can illuminate at TM polarization, where the zeroth order of the beam is not included in the light path of the probing spot. Aperture 2 was assisted by mounted iris (ID8, Thorlabs), which was used to avoid influence from the zeroth-order beam. Mirrors 2, 4, and 5 were assisted by a removable bracket using an indexing mount (NX1N/M, Thorlabs), which was placed to determine the position of the sample using widefield illumination. Especially, the same objective, excitation wavelength, and fluorescent filter were used both in the widefield and HiFi-FM.

The proposed HiFi-FM used scanning mechanisms via the point-by-point manner and capture rates of up to 10 frames/s. The modulated illumination beam in the Fourier plane of the microscope objective (100X/1.45 Oil, Apo Lambda, Nikon) and the sample were scanned by piezo ceramic stage (P-545.xR8S Piezo, PI) and controller (V31XYZ, Prior Scientific). By means of probing spot arrangement for raster scanning operation, a stack image of dimension 2.85 μm×2.85 μm was recorded by a scientific complementary metal-oxide semiconductor (sCMOS, Zyla 4.2 plus, Andor) with an image resolution of 150×150 pixels in 120 ms as opposed to an acquisition time of 50ms for piezo ceramic stage excitation. Although somewhat oversampled, we chose the above parameter to meet the Nyquist-Shannon sampling limit with 20 nm and reduce the acquisition time to ensure the performance of imaging. The fluorescence imaging emission filter (MF510-42, Thorlabs) with the bandwidth of 42 nm and a tube lens (TTL200-A, Thorlabs) with the focal length of 200 mm were placed in front of the sCMOS.

The scanning process with high-NA termed by the perfect focus system (PFS), which was effectively ensured the fluorescent sample well in the experiment with 21 µm×21 µm FOV. A detailed sketch of the experimental setup is presented in Fig. S1 of Supplement 1.

## 4. Experimental Results

### 4.1. Fluorescence beads imaging: resolution up to λ/8

To quantify the resolution enhancement of HiFi-FM, we have tested the resolution using fluorescence beads. With this type of sample, the key metric is which size of the beads can be distinguished and the intensity distribution can be regarded as the sparse sample on the dark-background of Eq. (5). The resolution of HiFi-FM image is significantly improved compared with widefield image with the NA=1.45, particularly in Figs. 5a and 5b white boxed regions. The beads are fluorescence-labeled by an emission wavelength of 518 nm and the average FWHM of the beads is 50 nm, with the input laser beam with the wavelength of 488 nm. Owing to the size of the fluorescence bead is negligible compared with the diffraction limit of 218 nm, the FWHM of a single bead can be considered as the resolution of the system. The enlarged white boxed region of image in Fig. 5b has indicated that two adjacent fluorescence beads can be clearly resolved in HiFi-FM, while in the widefield image shown in Fig. 5a, they are blurred and unresolvable. It is anticipated that HiFi-FM allows reconstruction resolution up to λ/8.

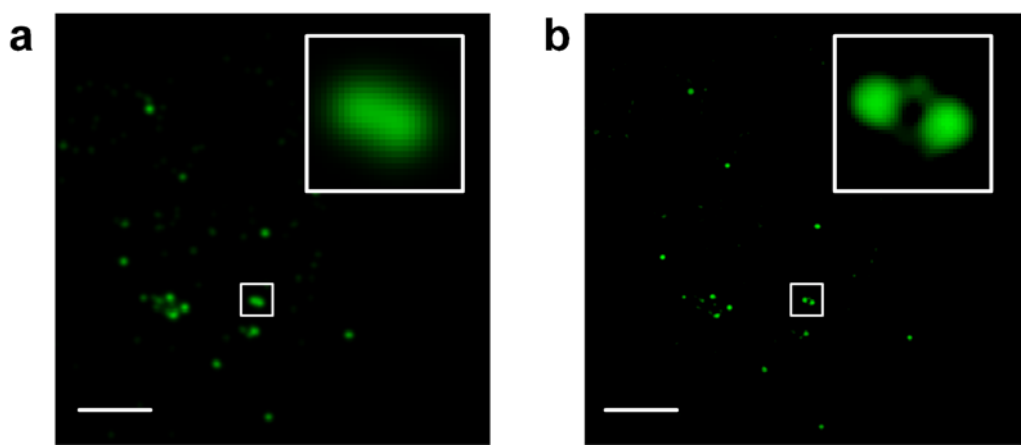

**Figure 5. Imaging of fluorescent beads by HiFi-FM with a resolution up to λ/8. a** Widefield image of fluorescence beads and enlarged white boxed region (upper right). **b** HiFi-FM image of fluorescence beads and enlarged white boxed region (upper right). Objective lens: 100×/1.45 NA. Scale bar: 1µm.

### 4.2. Tubulins of HEK 293T imaging: 21 µm×21 µm FOV with 0.13λ resolution

We have also opted to study morphologically and ultrastructurally distinct cells of human embryonic kidney 293T (HEK 293T) as a practical application of HiFi microscopy at λ/8 resolution. The distribution of intracellular structures can be regarded as an unknown complex distribution and the potential for degradation in image fidelity due to the convolutional effect of sidelobes in the illumination pattern should be considered. In fluorescence biological samples, the ground truth images are unknown so we cannot use contrast and SSIM as the measures of the fidelity of images. Fortunately, HiFi algorithm has been proven not sensitive to the reconstruction parameters and can be used to generate HiFi image, as shown in Fig. 3b. The tubulins of 293T cell are labeled with Alexa 488 which has maximal emission at a wavelength of 518 nm. Figures 6a-6c show tubulins of HEK 293T cell image acquired by widefield, NoFi, and HiFi. NoFi and HiFi images have been shown in Figs. 6b and 6c, and the image of NoFi is integrally stronger than other images owing to without considering the influence of the illumination sidelobes. The differences between HiFi and NoFi have been analyzed, and the intensity of $\lg[I(b)/I(c)]$ is shown in Fig. 6d, where $I(b)$ and $I(c)$ are respectively the intensity of Figs. 6b and 6c. Raster scanning images for tubulins of HEK 293T are shown in Fig. S2 of Supplement 1. We have chosen the solid region and dotted region with the zoomed-in image detailing presents the resolution of HEK 293T. The same sample, objective, excitation wavelength, and fluorescent filter have been used. Compared with that of the widefield image, the resolution of tubulins of HEK 293T in HiFi-FM is clearly improved. Quantitatively speaking, intensity profiles of both HiFi-FM and widefield images along the selected line using dotted line L1 and L2 are plotted in Figs. 6e and 6f. The tubulins of the 69 nm (~0.13λ) resolution with 21 μm×21 μm field of view can be acquired. Visualization 1 presents experimentally acquired images as a stack for the observed intensity to reconstruct the HiFi-FM image.

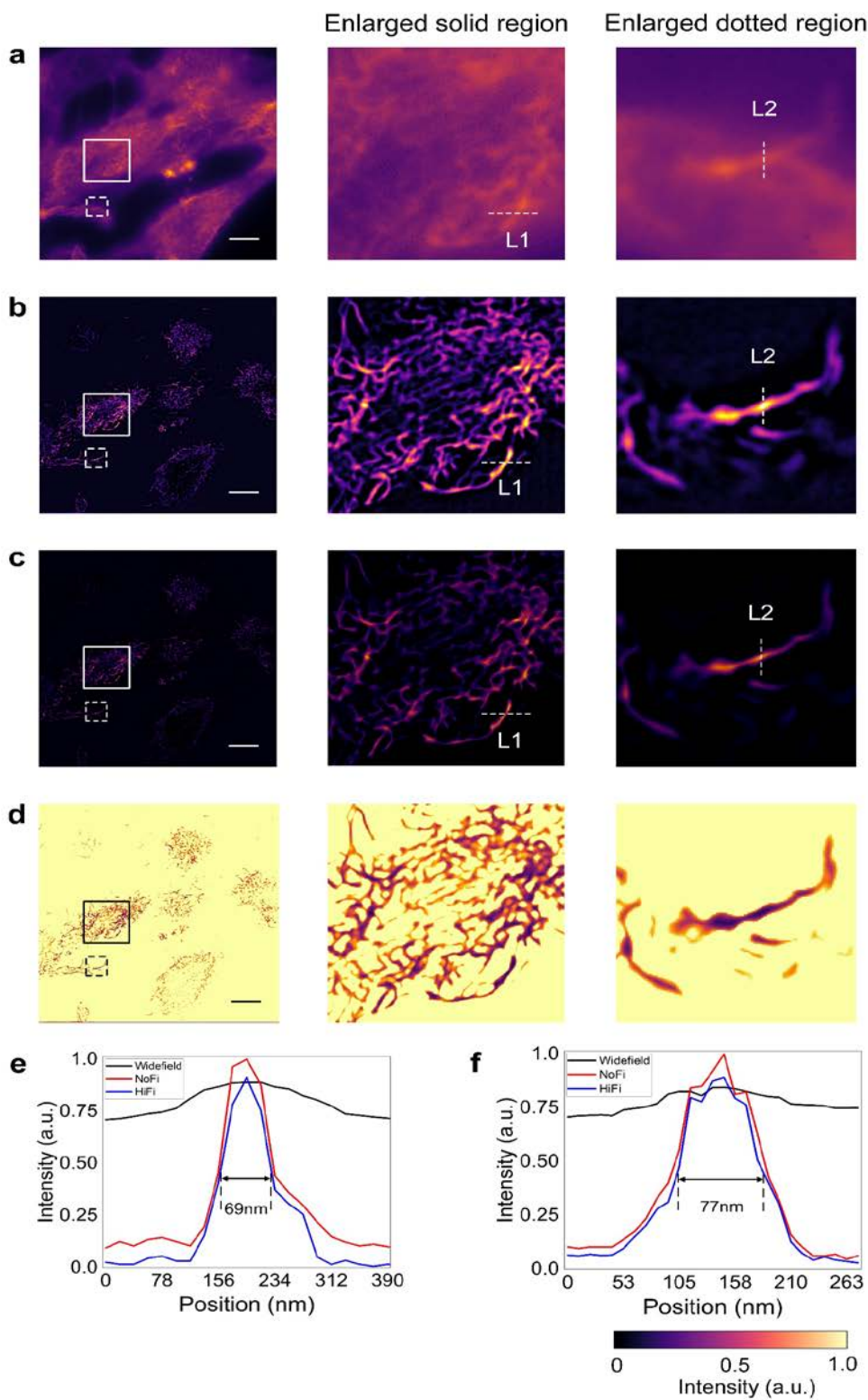

**Figure 6. Imaging of tubulins of HEK 293T with 21 µm×21 µm field of view. a** Widefield image of tubulins of HEK 293T and enlarged white solid and dotted boxed region. **b** NoFi image of tubulins of HEK 293T and enlarged white solid and dotted boxed region. **c** HiFi image of tubulins of HEK 293T and enlarged white solid and dotted boxed region. **d** The intensity by lg[*I*(b)/*I*(c)], where *I*(b) and *I*(c) are respectively the intensity of **b** and **c**. **e** Intensity profiles of tubulins of HEK 293T for white dotted line L1 on the selected line in **a-c**. **f** Intensity profiles of tubulins of HEK 293T for white dotted line L2 on the selected line in **a-c**. Scale bar: 2.5 µm. See Visualization 1.

## 5. Discussion and Conclusion

We have experimentally demonstrated that HiFi-FM allows reconstruction of spatially complex samples up to 69 nm (~0.13λ) resolution with the diffraction efficiency up to 3.76% while the HiFi result can be guaranteed. The idea of constrained super-resolution spots has been exploited for the first time to produce a sharp probing spot with the resolution of ~1/8 of the incident wavelength. By assisting the prior-knowledge, the super-resolution images with HiFi have been realized by proposed reconstruction algorithm.

The focused spot, which in principle has no physical limits on resolution, could be universally used for imaging at any wavelength. However, efficiency enhanced and resolution improved for super-resolution imaging is a shared concern in super-resolution imaging [24, 41]. With the use of diffractive optics assisted HiFi-FM, the probing spot with constrained spots is not circularly symmetric. Owing to the square target field in the iterative process, we do not use the optimization routine in polar coordinates while without the limitation of binary structure on the optimal solution. In addition, the constrained spots scheme precisely improves the resolution and efficiency of the probing spot.

To optimize DOE performance, tradeoffs are generally existing. There could be various other designs available with changes in the searching criterion. The diffraction efficiency of the probing spot smaller than 3.76% is acceptable, one would find many results demonstrating resolution higher than 65.2 nm. In our calculation and experiment, we choose the spot size with the acceptable efficiency to ensure a dynamic response quality for tubulins of HEK 293T imaging with 21μm×21μm FOV. Higher resolution may be realized with relaxed acceptance criterion of the maximum intensity. Inevitably, the unanticipated sidelobes and speckles still exist, which will influence the observed intensity. The used region of $P_{\text{illu}}(i, j)$ in HiFi image reconstruction algorithm should be increased to reduce the influence of such sidelobes and speckles. In the future, the preset area around the sample enables more prior-knowledge for image reconstruction. High-fidelity far-field super-resolution microscopy is ready to be implemented in a variety of fluorescent and label-free various samples.

## Supporting Information
Supporting Information is available from the Wiley Online Library or from the author.

## Acknowledgements
We acknowledge Dr. Mingjun Jiang at the Department of Biomedical Engineering, Peking University, China for her helpful assistance during sample preparation. We acknowledge support from the National Natural Science Foundation of China (Grants No. 62075112).

**Conflict of Interest**
The authors declare that there are no conflicts of interest related to this paper.

**Data Availability Statement**
The data that support the findings of this study are available from the corresponding author upon reasonable request.

References (Full information for editor and reviewers)

# Supporting Information

## High-fidelity far-field microscopy at λ/8 resolution

*Ning Xu, Guoxuan Liu, and Qiaofeng Tan\**

### Supplementary Note

| | |
|---|---|
| **Supplementary Note 1** | High-fidelity image reconstruction |
| **Supplementary Note 2** | Sample preparation |
| **Supplementary Note 3** | Visualization 1 |
| **Supplementary Figure 1** | Experimental setup of HiFi-FM |
| **Supplementary Figure 2** | Raster scanning images for tubulins of HEK 293T |
| **Supplementary Figure 3** | Flow chart of high-fidelity image reconstruction procedure |
| **Supplementary Figure 4** | Area distributions |
| **Supplementary Figure 5** | Quantitative characterization of the fidelity of the sample |

## 1. High-fidelity Image Reconstruction

To reconstruct the high-fidelity (HiFi) image with super-resolution, we applied an iterative reconstruction algorithm originated from the Richardson-Lucy deconvolution algorithm [1, 2]. The outputs of the deconvolution algorithm are the object reconstructions. This deconvolution algorithm is given by

$$e_{k+1} = e_k \cdot \left( \frac{T_{\text{in}}}{e_k \otimes p_{\text{imag}}} \otimes p_{\text{imag}}^{\text{T}} \right), \tag{S1}$$

where $e_k$ is the $k$th estimate of the object, $T_{\text{in}}$ is the input image that is also equal to $e_0$, $p_{\text{imag}}$ is the PSF of the imaging system, and $p_{\text{imag}}^{\text{T}}$ is the transpose of $p_{\text{imag}}$. The brief reconstruction process of the HiFi image reconstruction is introduced in Fig. S3.

The PSFs with lateral FWHM of the illumination system and the imaging system can be calibrated from measurements on 50 nm fluorescent beads. During data acquisition, widefield images and raw images are captured. The dark region in the sample is selected as the prior-knowledge to serve as the criterion of convergence. For the resolution test target (known the ground-truth (GT) information), the intensities of horizontal and vertical directions in dark regions are closer to zero, the results are closer to fidelity. For the fluorescent sample (unknown the GT information), the dark regions are selected by widefield and non-fidelity (NoFi) images, i.e. the intensities of both illuminations are reaching zero, the selected prior-knowledge area is shown in Fig. S4b (the area beyond the white dotted line).

After selecting the dark areas, the transmittance values of $T_p$ need to select at the positions where the probing spot does not illuminate the sample while the sidelobes do during the scanning, as shown in Fig. S4c. Because $T(m, n)$ should be zero when the probing spot is in the dark region, however the intensity values $I(m, n)$ are not zero due to the effect of convolution. Such a transmittance $T_p$ also can be considered as the prior-knowledge. In our system, about 10% area of biological samples is suitable to obtain good performance with a short time. The initial estimated value can be expressed as Eq. (12). In each iteration, the known-transmittance areas have been calculated as criteria. To simplicity the calculation while reducing the sensitivity of parameters, the values of know-transmittance area are set to zeroes, as shown in Fig. S4. We performed 100 times iteration to ensure the convergence with $\alpha$=0.27, $C_1$ and $C_2$ are equally chosen as $C_1$= $C_2$=0.03 to avoid division by a small denominator.

## 2. Sample Preparation

The prepared Lumisphere monodisperse microsphere solution is a fluorescent microsphere with a radius of 50nm (7-3-0010, Basel). Take 1mL of fluorescent microsphere solution and drop it on a 0.17mm cover glass and washed 4 times with alcohol buffer before imaging.

Hek293-T cells were cultured in DMEM with 10% fetal bovine serum (FBS, Gibco) and 1% penicillin/streptomycin. For immunofluorescent staining, cells were washed 2 times with pre-warmed Dulbecco's phosphate-buffered saline (DPBS, Gibco) and fixed in 4% paraformaldehyde for 10 min. Permeabilize the cells with 0.1% Triton X -100 in PBS for 15 min and block with 1% bovine serum albumin (BSA) for 1 hour. Alexa Fluor Dyes (A12379, Invitrogen) were purchased from ThermoFisher and prepared as stock solutions following the protocols provided. Staining was conducted in Hank's balanced salt solution (HBSS). The optimized final working concentrations for Alexa Fluor™ 488 were 0.86 μg/mL ($2.5^x$). Incubate the cells with prepared stain reagents in the dark for 1 h and washed with HBSS 2 times before imaging according to manufacturer's recommendations. After dyeing the Hek 293T cells, applying 3 drops (~100μL) of the glass antifade mountant (P36982, Invitrogen) are directly onto the specimen. Place the mounted sample on a flat, dry surface for 48 hours at room temperature in the dark.

## 3. Supplementary Note for Visualization 1:

Visualization 1 shows a probing super-resolution spot focused on tubulins of HEK 293T with piezo ceramic stage. The objective with NA equal to 1.45 and the excitation wavelength of 488 nm is used.

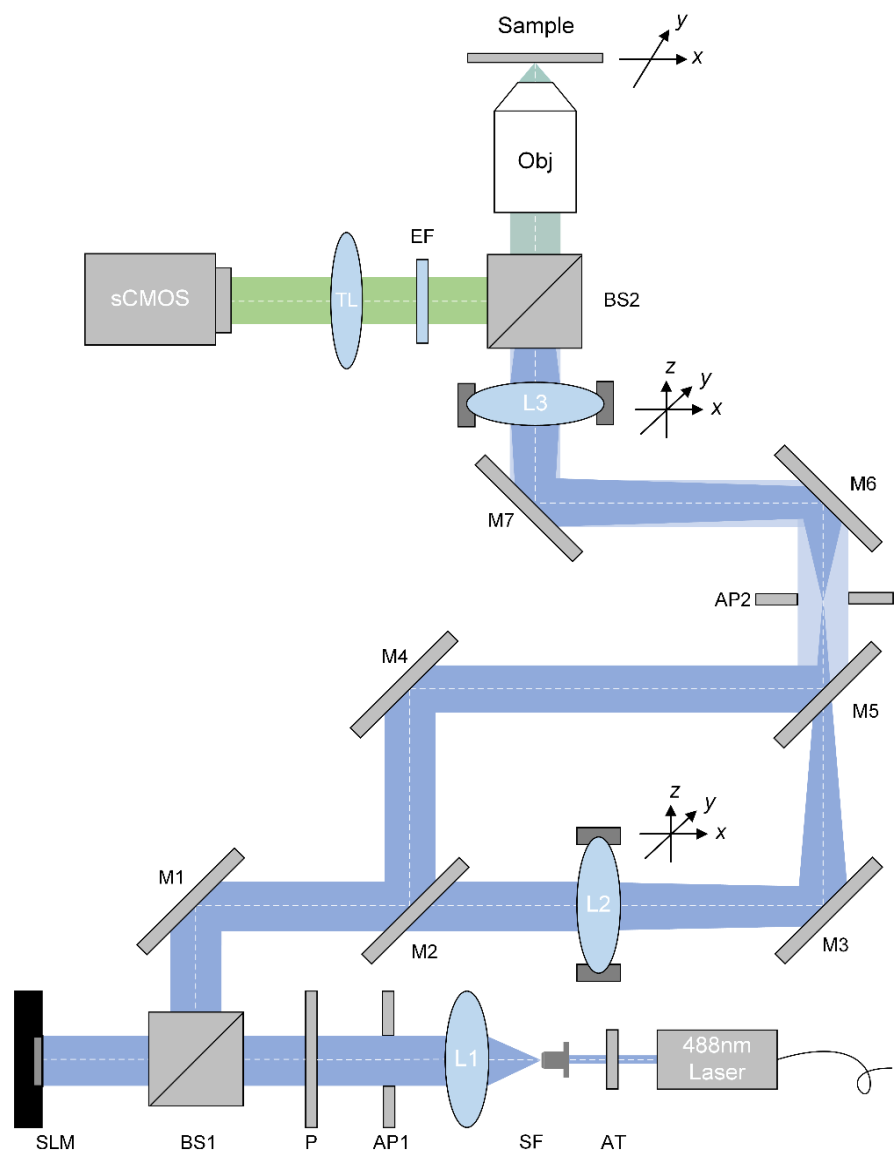

**Fig. S1. Experimental setup of high-fidelity far-field microscopy (HiFi-FM).** The experimental setup consists of two main system: illumination system (Blue ray) and imaging system (Green ray). A collimated laser beam illuminates the spatial light modulator (SLM) via a beam splitter. A designed phase uploaded on the SLM is diffracted in the sample placed on a piezo ceramic stage through a microscope objective. Sample scanned by the stage is captured onto the sCMOS through tube lens and emission filter. AT, attenuator; SF, spatial filter; AP, aperture; P, polarizer; BS, beam splitter; SLM, spatial light modulator; M, mirror; L, lens; Obj, objective; EF, emission filter; TL, tube lens.

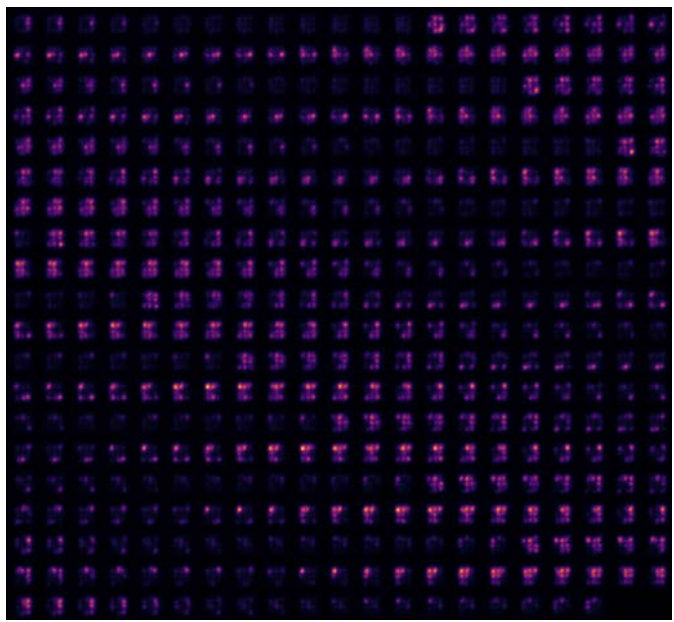

**Fig. S2. Raster scanning images for tubulins of HEK 293T after intensity correction.** We captured each image with an image resolution of 150×150 pixels in 120 ms as opposed to an acquisition time of 50 ms for piezo ceramic stage excitation for tubulins of HEK 293T using raster scanning method. The step with 20 nm is used to reduce the acquisition time with the magnification of 375× for imaging. The intensities of point-by-point scanning data have been corrected by Note 1 in supplementary materials.

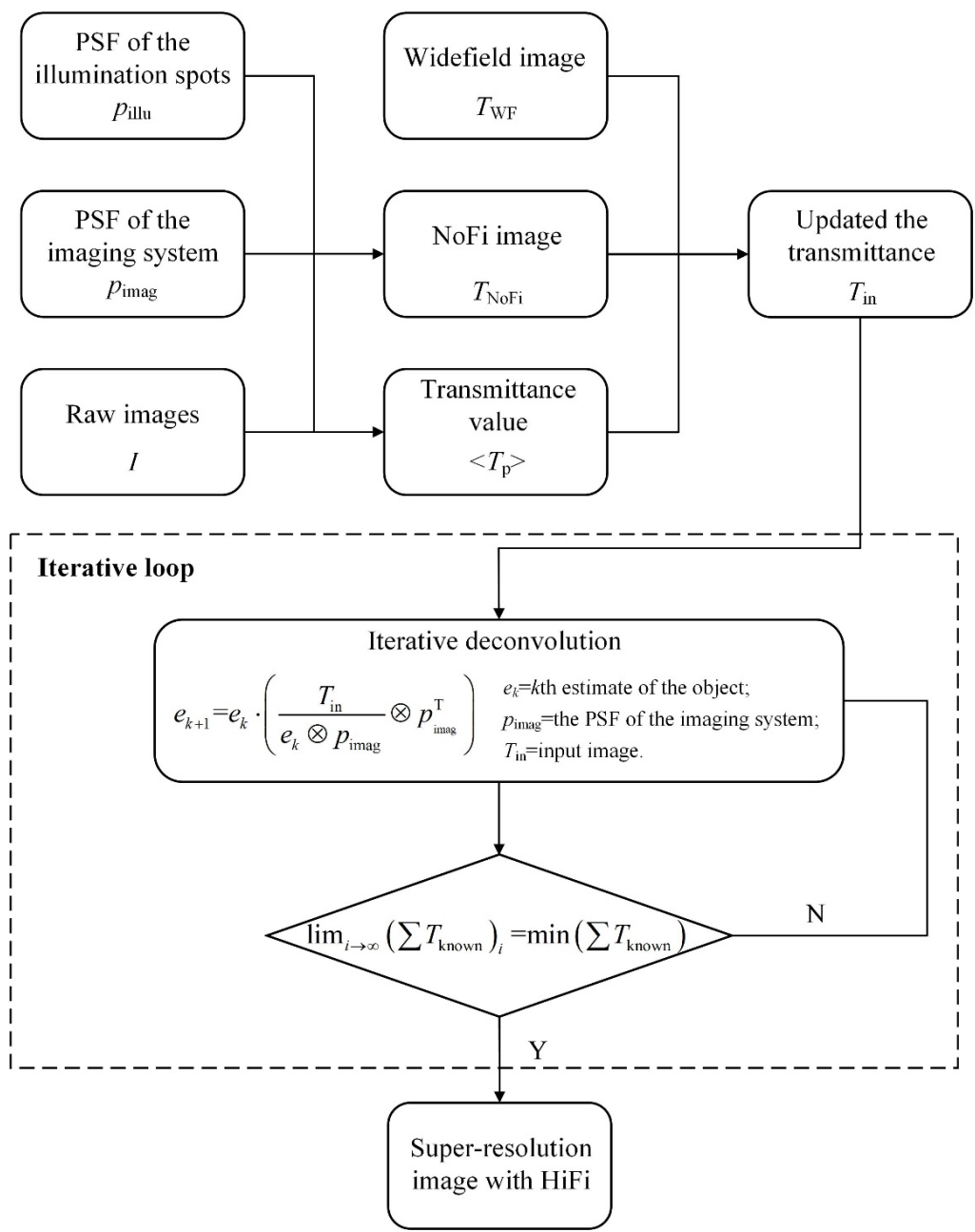

**Fig. S3.** Flow chart of HiFi image reconstruction procedure.

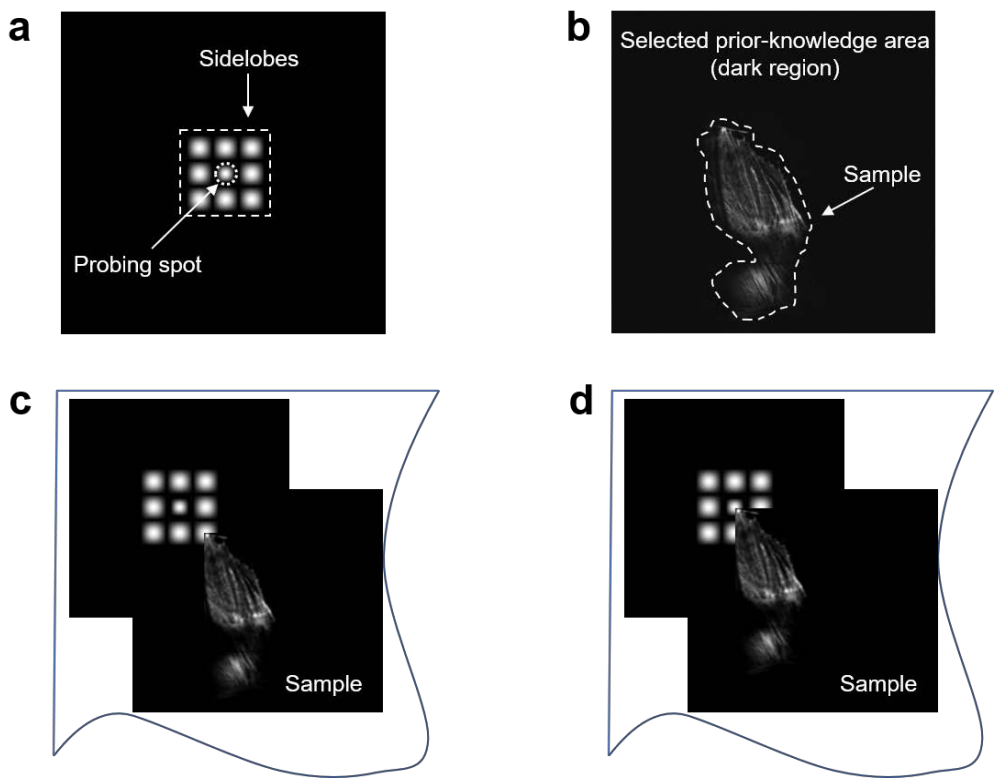

**Fig. S4. Area distributions. a** The probing spot area. **b** Prior-knowledge area in the sample plane with the transmittance value near to zero. **c** The transmittance $T_p$ at the positions where the probing spot does not illuminate the sample while the sidelobes do during the scanning, and **d** The probing spot illuminates the sample.

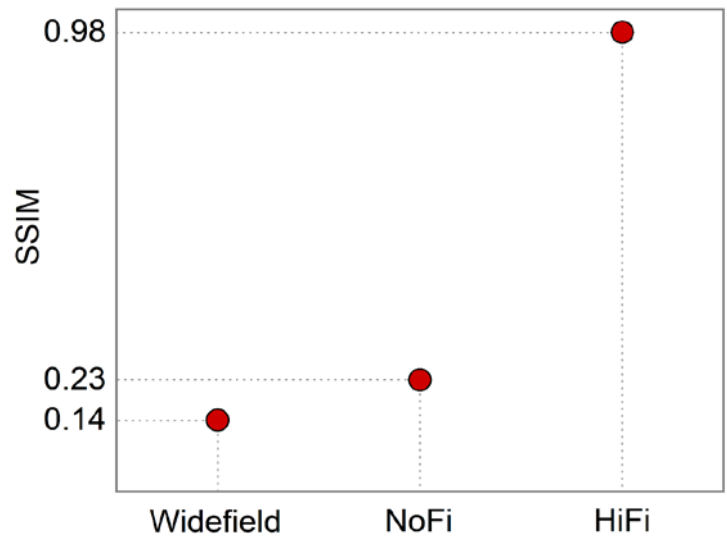

**Fig. S5.** Quantitative characterization of the fidelity of the positive resolution test target reconstruction by structural similarity index measure (SSIM) versus widefield, NoFi, and HiFi.